\documentclass[twocolumn,trackchanges,appendixfloats]{aastex}
\newcommand{\N}[1]{$N$(#1)}

\begin{document}

\title{A Surface Density Perturbation in the TW~Hydrae Disk\\at 95~au Traced by Molecular Emission}

\author{R. Teague\altaffilmark{1,2},
D. Semenov\altaffilmark{1},
U. Gorti\altaffilmark{3,4},
S. Guilloteau\altaffilmark{5,6},
Th. Henning\altaffilmark{1},
T. Birnstiel\altaffilmark{1}, \\
A. Dutrey\altaffilmark{5,6},
R. van Boekel\altaffilmark{1}
\& E. Chapillon\altaffilmark{5,6,7}
}

\altaffiltext{1}{Max-Planck-Institut f\"{u}r Astronomie, K\"{o}nigstuhl 17, 69117 Heidelberg, Germany}
\altaffiltext{2}{teague@mpia.de}
\altaffiltext{3}{NASA Ames Research Center, Moffett Field, CA 94035, USA}
\altaffiltext{4}{SETI Institute, Mountain View, CA 94043, USA}
\altaffiltext{5}{University Bordeaux, LAB, UMR 5804, 33270 Floirac, France}
\altaffiltext{6}{CNRS, LAB, UMR 5804, 33270 Floirac, France}
\altaffiltext{7}{IRAM, 300 rue de la Piscine, Domaine Universitaire, F-38406 Saint Martin d'H\'{e}res, France}

\begin{abstract}
We present ALMA Cycle~2 observations at $0.5\arcsec$ resolution of TW~Hya of CS $J=5-4$ emission. The radial profile of the integrated line emission displays oscillatory features outwards of $1.5\arcsec$ ($\approx 90$~au). A dip-like feature at $1.6\arcsec$ is coincident in location, depth and width with features observed in dust scattered light at near-infrared wavelengths. Using a thermochemical model indicative of TW~Hya, gas-grain chemical modelling and non-LTE radiative transfer, we demonstrate that such a feature can be reproduced with a surface density depression, consistent with the modelling performed for scattered light observations of TW~Hya. We further demonstrate that a gap in the dust distribution and dust opacity only cannot reproduce the observed CS feature. The outer enhancement at $3.1\arcsec$ is identified as a region of intensified desorption due to enhanced penetration of the interstellar FUV radiation at the exponential edge of the disk surface density, which
intensifies the photochemical processing of gas and ices.
\end{abstract}

\maketitle

\section{Introduction}
\label{sec:introduction}

Protoplanetary disks are sites of active planet formation and it is in the dense midplanes of these disks where grain growth seeds planetesimal formation. Once a planet has grown to a certain mass, it will begin interacting with the disk and sculpt the density structure \citep{Kley_Nelson_2012, Turner_PPVI_2014}. The hall marks of ongoing planet formation and planet-disk interactions are frequently observed such as spirals and gaps in scattered light \citep{Debes_ea_2013, Benisty_ea_2015, Wagner_ea_2015} or rings and dust traps seen in thermal mm-dust continuum \citep{vanderMarel_ea_2013, ALMA_ea_2015, Andrews_Wilner_ea_2016}.

While embedded planets are an attractive mechanism for producing these features, other physical mechanisms unrelated to planets have been shown to produce similar features, such as gravitational instabilities, dead zones of the magneto-rotational instability (MRI), snowlines of volatile species or dust evolution \citep{Flock_ea_2015, Pohl_ea_2015, Zhang_ea_2015, Birnstiel_ea_2015}. To connect these disparate regions requires studies of molecular line emission which traces the gas between the areas probed by thermal continuum or scattered light.

\begin{figure*}
\centering
\includegraphics[]{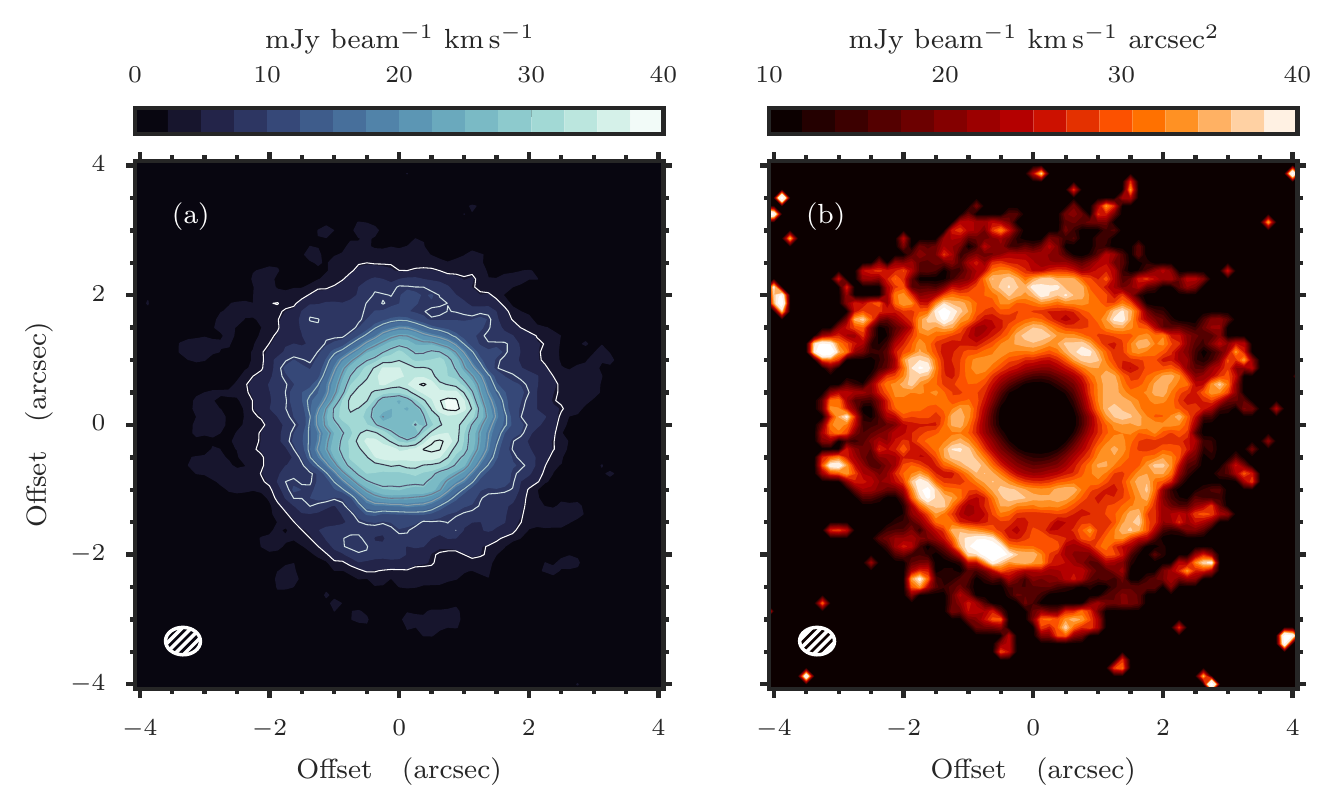}
\caption{(a) A zeroth moment map of CS $J=5-4$ in TW~Hya. The contours are in steps of $3\sigma$ where $\sigma = 5.35~{\rm mJy~beam^{-1}~km\,s^{-1}}$ and the beam, with size $0.42\arcsec \times 0.54\arcsec$, is shown in the bottom left hand corner. (b) The zeroth moment map from (a) with an $r^2$ scaling applied to highlight the features in the outer disk. Such a scaling is used only to highlight the outer disk and is not physically motivated. A dip is clearly seen at $r \approx 1.6\arcsec$, coincident with the dip in scattered light emission \citep{Debes_ea_2013,vanBoekel_ea_2016}. \label{fig:observations_zeroth}}
\end{figure*}

TW~Hya, as the nearest protoplanetary disk to us at $59\pm1$~pc \citep{Gaia_2016}, is the most well studied disk across a huge wavelength range and is the ideal target for such a study. The disk displays a variety of substructures and radial morphologies in scattered light \citep{Debes_ea_2013, Akiyama_ea_2015, Rapson_Kastner_ea_2015, vanBoekel_ea_2016}, mm-dust continuum \citep{Andrews_ea_2012, Menu_ea_2014, Andrews_Wilner_ea_2016, Tsukagoshi_ea_2016} and molecular line emission \citep{Qi_ea_2013, Kastner_ea_2015, Nomura_ea_2016, Schwarz_ea_2016, Bergin_ea_2016}, with many physical and chemical processes being invoked to explain these striking features.

In this paper we present ALMA Cycle~2 observations of CS $J=5-4$ line emission whose radial profile shows a shallow dip coincident with the $\sim 90$~au dust gap seen in the scattered light \citep{Debes_ea_2013, Rapson_Kastner_ea_2015, vanBoekel_ea_2016}. Section~\ref{sec:observations} describes the observations and the observed features while in Section~\ref{sec:modelling} we follow the modelling carried out by \citet{vanBoekel_ea_2016} and argue that these features are the product of a depression in the surface density of the gas rather than a chemical effect. Our findings are discussed in more detail, including their power as complimentary studies for future high-resolution observations, in Section~\ref{sec:discussion}. A summary follows in Section~\ref{sec:conclusion}.

\section{Observations} 
\label{sec:observations}

\subsection{Data Reduction}
\label{sec:data_reduction}

The observations targeted CO $J=2-1$ at 230.538~GHz, CN $N=2-1$ at 226.875~GHz and CS $J=5-4$ at 244.936 GHz and were originally presented in \citet{Teague_ea_2016}. However, we provide a brief description of the data here and focus only on the CS $J=5-4$ emission. 

The ALMA Cycle 2 observations, project 2013.1.00387.S, were performed on May 13, 2015. Data were calibrated using the standard ALMA calibration script in the \texttt{CASA} software package\footnote{\url{http://casa.nrao.edu/}}. The absolute flux calibration was estimated using Ganymede. The derived flux for our amplitude and phase calibrator, J1037-2934, was 0.72~Jy at 228~GHz at the time of the observations, with a spectral index $\alpha = -0.54$, while the ALMA flux archive indicated a flux of $0.72 \pm 0.05$ Jy between April 14$^{\rm th}$ and April 25$^{\rm th}$. We hence estimate that the calibration uncertainty is about 7\%.

Figure~\ref{fig:observations_zeroth} shows the continuum subtracted zeroth moment of the CS $J=(5-4)$ line emission in panel (a).  No strong azimuthal structure is observed to be significant when considering the noise. The integrated flux of the line was 1.24~Jy~beam$^{-1}$~km\,s$^{-1}$, calculated from an $8\arcsec \times 8\arcsec$ box centred on the source. Panel (b) shows the same moment map, with an $r^2$ scaling applied in order to highlight the outer regions of the disk. This scaling is applied to scattered light data to account for the drop in stellar flux at a given radial position, however for molecular line emission, such is the case here, this is not physically motivated and applied purely to highlight the outer regions of the disk.

\begin{figure}
\includegraphics[width=\columnwidth]{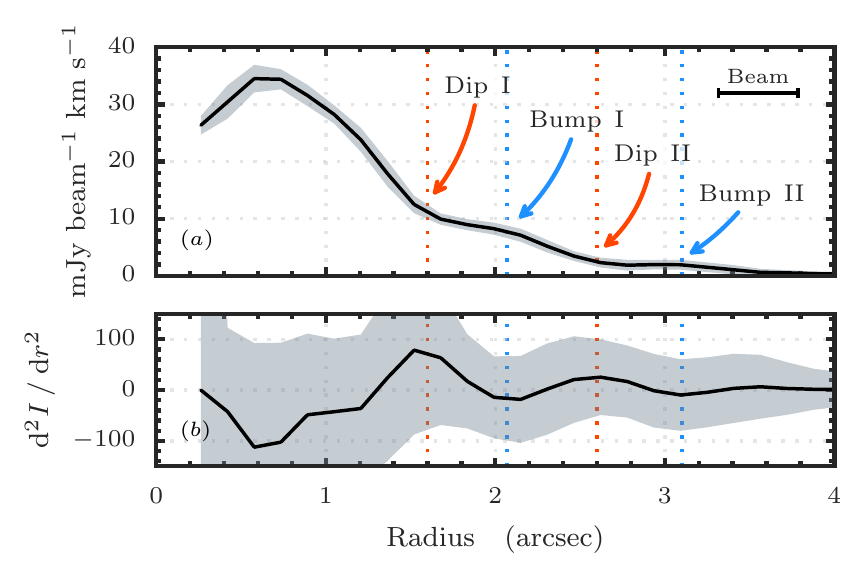}
\caption{Top: radial profile of the CS $J=5-4$ emission from Fig.~\ref{fig:observations_zeroth}a. Annotated are the potential dip and enhancement locations. The beam FWHM is shown in the top right corner. Bottom: second derivative of the radial profile highlighting the features. The gray regions show $1\sigma$ standard deviation of each azimuthal bin}. \label{fig:observations_radialprofile}
\end{figure}

As no azimuthal structure is seen in the zeroth moment maps, nor suggested in previous observations, these data were azimuthally averaged assuming an inclination of $7\degr$ \citep{Qi_ea_2004} and position angle $240\degr$ \citep{Teague_ea_2016} to yield a radial intensity profile, shown in Fig.~\ref{fig:observations_radialprofile}. The solid line shows zeroth moment profile, while the dotted line shows the $r^2$ scaling (panels (a) and (b) from Fig.~\ref{fig:observations_zeroth} respectively). Gray shading around the lines demonstrate the 1\,$\sigma$ noise in the image. This does not take into account the estimated flux calibration uncertainty of 7~\% which would be a global rescaling of the profile.

\subsection{Radial Features}
\label{sec:radial_features}

The peak of the emission is offset from the centre at $\approx 1.2\arcsec$. Within this radius the continuum emission dominates the weaker CS emission, thus continuum subtraction may affect the assumed inner radii of CS. Outside of this, the emission falls off and is detected to $\sim 3.7\arcsec$ \citep{Teague_ea_2016}, comparable to that of CO and scattered light \citep{Andrews_ea_2012, Debes_ea_2013}.

The integrated intensity emission profile shows oscillatory features between outwards of 1.5$\arcsec$, as noted on Fig.~\ref{fig:observations_radialprofile}. Two features are present, an inner and an outer feature, each consisting of a `dip' and a `bump' depending on the sign of the second derivative, as shown in the bottom panel of Fig~\ref{fig:observations_radialprofile}. As there is no baseline to make a comparison with, these features could be considered dips in emission at $1.6\arcsec$ and $2.6\arcsec$, enhancements at $2.2\arcsec$ and $3.2\arcsec$, or a combination of both.

As the mm~continuum has a sharp truncation at $\sim 60$~au \citep{Hogerheijde_Bekkers_ea_2016}, a comparison with the substructures observed by \citet{Andrews_Wilner_ea_2016} is impossible. Scattered light emission, however, extends out to $\sim 220$~au making for a better comparison \citep{Akiyama_ea_2015, Rapson_Kastner_ea_2015, vanBoekel_ea_2016}. The most prominent feature, `Dip I', at $1.6\arcsec$, or 95~au when assuming a distance of $d = 59$~pc \citep{Gaia_2016}\footnote{At the pre-Gaia distance of 54~pc \citep{vanLeeuwen_2007}, this would be 86~au.}, matches the location and apparent depth to the dip in scattered light emission previously observed with NICMOS on the Hubble Space Telescope \citep{Debes_ea_2013} and later confirmed with ground-based observations. Such a similarity is suggestive of a common origin for these two features, a scenario which is explored further in Section~\ref{sec:modelling}.

An alternative scenario is that the features are enhancements in emission. Similar outer enhancements in molecular emission have been observed in several disks, for example the secondary ring of DCO$^+$ in IM~Lup \citep{Oberg_ea_2015} and in CO isotopologes in AS~209 and the inner region of TW~Hya \citep{Huang_ea_2016, Schwarz_ea_2016}. These works have all invoked a secondary desorption front, either photo or thermal, which releases volatiles from the ices, resulting in enhanced emission locally. However, these scenarios require a change in the local grain properties to produce a change in either UV penetration or thermal structure \citep{Cleeves_2016}. This prerequisite is satisfied for all three cases where the enhancement is found near the edge of the mm-continuum, a proxy for the edge of the mm grains. For the two enhancements described here, `Bump I' and `Bump II', the edge of the millimetre grains is too far inwards to be the cause. However, `Bump II' is very close to the outer edge of the disk, and so may be due to enhanced desorption. This scenario is discussed in more detail in Section~\ref{sec:discussion}.

As much work has been done on secondary desorption fronts, in this paper we focus on the possibility of a feature induced by a surface density perturbation, as used in models of scattered light observations.

\section{Modelling Surface Density Perturbations}
\label{sec:modelling}

A commonly invoked mechanism to account for dips in scattered light emission is a perturbation in the total surface density, a method used by both \citet{Debes_ea_2013} and \citet{vanBoekel_ea_2016} to model the TW~Hya scattered light emission. In this section, we explore, using a model representative of TW~Hya the impact of such a surface density perturbation on the molecular emission of CS. 

As molecular emission is a product of both excitation and abundance effects, it is important to perform self-consistent modelling taking into account the impact of the surface density perturbations on the physical structure and the resulting change in chemical abundances. To do this, we use an advanced thermochemical model in conjunction with chemical modelling and non-LTE radiative transfer to demonstrate the effect of a surface density perturbation on the radial profiles of the molecular emission and distinguish between scenarios.

\subsection{Physical Structure}

\begin{figure*}
\plotone{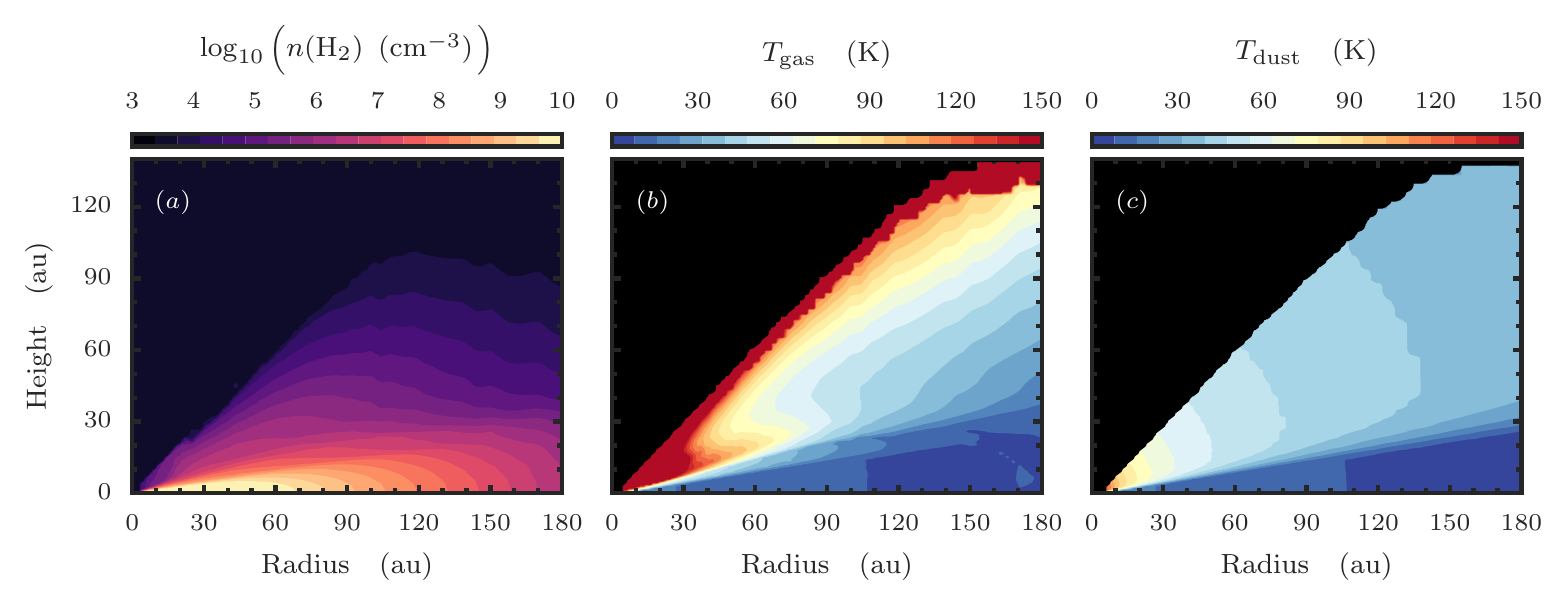}
\caption{Gas density (left), temperature (center) and dust temperature (right) for the fiducial model, Model A. \label{fig:physical_structure}}
\end{figure*}

For a baseline, we take the TW~Hya model of \citet{Gorti_ea_2011} as the fiducial model. The unperturbed surface density is given by,

\begin{equation}
\Sigma_0 = 500 \,\, r^{-0.7} \, \exp\left(-\frac{r^{1.3}}{100}\,\right) \quad {\rm g \, cm^{-2}},
\label{eq:surfacedensity}
\end{equation}

\noindent where $r$ is the radial distance in au, yielding a total gas mass of $M_{\rm disk} = 0.06~M_{\sun}$. The distribution was derived by fitting various gas spectra and dust continuum observed from the sub-millimetre to to optical wavelengths. The model was made to reproduce integrated intensities rather than spatially resolved observations. Thus, while the model may not fully reproduce spatially resolved observations, it is a model with temperatures and densities expected in TW~Hya. The model assumes a gas-to-dust ratio of 100, with an MRN-like grain size distribution, $n(a) \propto a^{-3.5}$, with maximum and minimum sizes of 1~mm and 0.09~$\micron$ respectively. The assumed relative abundance of polycyclic aromatic hydrocarbons (PAH) per hydrogen nuclei is $10^{-9}$. The unperturbed model, Model A, is shown in Fig.~\ref{fig:physical_structure} with the gas density, gas temperature and dust temperature in the panels, running left to right. The total mass of the unperturbed model is 0.05~$M_{\sun}$, consistent with the masses derived from HD measurements \citep{Bergin_ea_2013}.

The surface density is then perturbed with a Gaussian depression characterised by center $d_0$, width $\Delta d$ and depth $d$ such that the resulting surface density is given by,

\begin{equation}
\Sigma \, (r) = \Sigma_0 \, (r) \times \left(1 - d \cdot \exp \left[ -\frac{1}{2}\frac{(r - d_0)^2}{\Delta d^2} \right] \right).
\label{eq:perturbation}
\end{equation}

\noindent We consider three perturbations to the surface density: Model B, a perturbation in the total surface density as in \citet{Debes_ea_2013}, $d = 0.3$, $\Delta d = 20$~au and $d_0 = 80$~au; Model C, the outer perturbation from \citet{vanBoekel_ea_2016}, $d = 0.55$, $\Delta d = 15$~au and $d_0 = 80$~au; and Model D, a fully depleted gap, $d = 1$, $\Delta d = 9.45$~au and $d_0 = 80$~au. The widths have been chosen such that each perturbed model has the same amount of mass removed relative to the unperturbed Model A: $2.7 \times 10^{-3}~M_{\sun}$ ($2.8~M_{\rm Jup}$). This criteria for the width of the gap recovers the best-fit widths found with both Models B and C \citep{Debes_ea_2013,vanBoekel_ea_2016}.

Given that the reduction of emission in scattered light could be due to a local depletion of small grains, we consider an additional set of models where the perturbation is applied only to the dust surface density. This results in the perturbed regions possessing an enhanced gas-to-dust ratio relative to the fiducial 100:1. With these models it is possible to explore the sensitivity of the molecular emission to the local available grain surface.

\subsection{Chemistry}

With the prescribed surface densities covering between 3.9 and 200~au (TW~Hya is observed to have material as close in as 1~au, however this will not contribute to the low-energy molecular line emission concerned with here), the 1+1D disk physical structure was solved for self-consistently including heating and cooling processes following \citet{Gorti_ea_2011}. Gas and dust temperatures are treated independently allowing the gas temperatures in the upper, more strongly irradiated regions of the disk to deviate strongly from those of the dust. The stellar radiation was set-up assuming a central stellar mass of $0.7~M_{\sun}$, radius $1.1~R_{\sun}$ and effective temperature of 4200~K. This comprises of a total far-UV luminosity of $3\times10^{31}$~erg\,s$^{-1}$ and an X-ray spectrum that covers $0.1$--$10$~KeV and a total X-ray luminosity of $1.6\times10^{30}$~erg\,s$^{-1}$. The accretion rate is assumed to be $10^{-9}~\dot{M}_{\sun}$~yr$^{-1}$, as constrained by the X-ray and UV observations \citep[e.g.,][]{Brickhouse_ea_2012}.

\begin{figure*}
\centering
\includegraphics[]{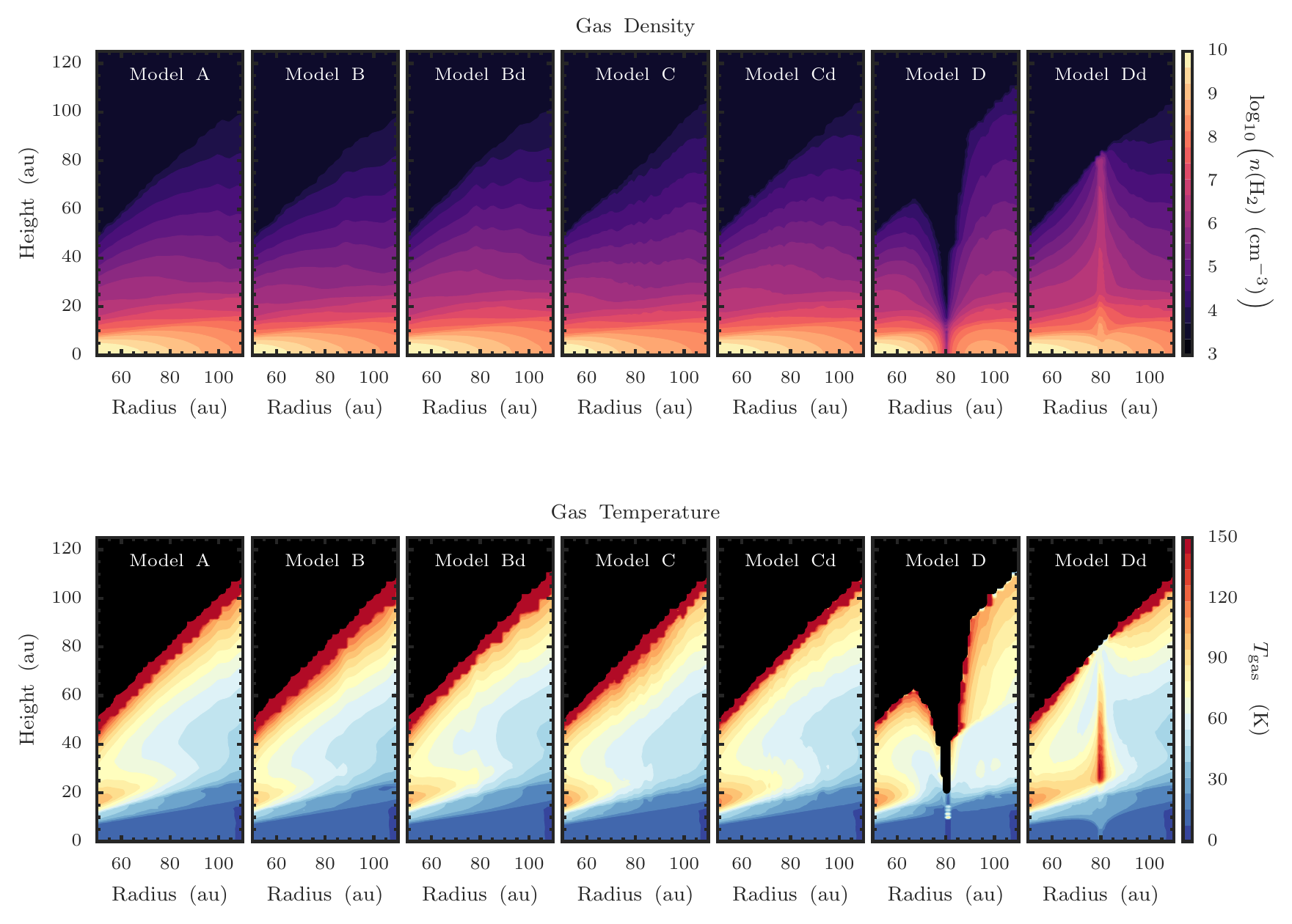}
\caption{Comparison of the $n({\rm H}_2)$, top, and $T_{\rm gas}$, bottom, structures for the seven models over the gap location, $80 \pm 30$~au. The physical structure outside this radial region is comparable for all seven molecules. As the dust temperature structure is indistinguishable between the models and so not shown. \label{fig:density_temperature_slices}}
\end{figure*}

The resulting physical structures were used as the basis for the chemical modelling with the 1+1D code \texttt{ALCHEMIC} \citep{Semenov_ea_2010}, employing a full gas-grain network with deuterium fractionation, including up to triply-deuterated species \citep{Albertsson_ea_2014a}, and assuming uniformly-sized amorphous olivine grains with the radius $a_d = 7\micron$. This grain size choice results in a grain surface per cell equivalent to that of the size ensemble used in the physical modelling. 

In the chemical model, X-ray and UV photons and cosmic ray particles (CRPs) are the sources of ionizing radiation. The standard CRP ionization rate of $1.3 \times 10^{-17}$~s$^{-1}$ is adopted throughout the entire disk. The FUV penetration is computed in an 1+1D approximation as in \citet{Semenov_Wiebe_2011}, using the visual extinction $A_{\rm V}^{\star}$ in the direction towards the central star for the stellar FUV component, and the visual extinction $A_{\rm V}^{\rm IS}$ in the vertical direction for the interstellar (IS) FUV component. To compute visual extinction $A_{\rm V}$, we used the following approximate scaling relation:

\begin{equation}
A_{\rm V} \, ({\rm mag}) = \frac{N({\rm H}) \, ({\rm cm^{-2}})}{1.59 \times 10^{21}} 
\frac{0.1}{a_d \, (\micron)} \frac{m_{d/g}}{10^{-2}},
\end{equation}

\noindent where $a_d$ is the grain radius in $\micron$ and $m_{d/g}$ is the dust-to-gas mass ratio. The unattenuated stellar FUV intensity at 100~au is $\chi_{\star} = 410\chi_0$ as expressed in the \citet{Draine_1978} FUV field units. The unattenuated interstellar FUV intensity $\chi_{\rm IS} = 1\chi_0$. Then, the FUV flux at a given disk location could be roughly computed as

\begin{equation}
I_{\rm UV} =  \chi_{\star} \, e^{-A_{\rm V}^{\star} \,/\, 1.086} + \chi_{\rm IS} \, e^{-A_{\rm V}^{\rm IS} \,/\, 1.086}.
\end{equation}

\noindent To compute the X-ray ionization rate, we used the parametrization of \citet{Bai_Goodman09}, as presented in Eq.~(37) from \citet{Armitage_SF_2015} for the average X-ray photon energy of 3~keV.

\floattable
\begin{deluxetable}{cccc}
\tabletypesize{\footnotesize}
\tablecolumns{4}
\tablenum{1}
\tablecaption{Initial Chemical Abundances \label{tab:initial_abundances}}
\tablehead{
	\colhead{Species} & \colhead{Abundance\,\tablenotemark{a}} & 
    \colhead{Species} & \colhead{Abundance\,\tablenotemark{a}} \\ 
    \colhead{} & \colhead{($n$(X) / $n$(H))} & \colhead{} & \colhead{($n$(X) / $n$(H))}
}
\startdata
ortho-H$_2$ & $3.75\;(-1)\phn$ & S	& $9.14\;(-8)\phn$ 	\\
para-H$_2$  & $1.13\;(-1)\phn$ & Si & $9.74\;(-9)\phn$ 	\\
He 			& $9.75\;(-2)\phn$ & Fe & $2.74\;(-9)\phn$ 	\\  
O  			& $1.80\;(-4)\phn$ & Na & $2.25\;(-9)\phn$ 	\\
C  			& $7.86\;(-5)\phn$ & Mg & $1.09\;(-8)\phn$ 	\\  
N  			& $2.47\;(-5)\phn$ & Cl & $1.00\;(-9)\phn$ 	\\
HD 			& $1.55\;(-5)\phn$ & P  & $2.16\;(-10)$ 	\\
\enddata
\tablenotetext{a}{Read $a(b)$ as $a \times 10^{-b}$.}
\end{deluxetable}

A `low metals', mainly atomic set of initial abundances was used \citep{Lee_ea_1998, Semenov_ea_2010}. Table~\ref{tab:initial_abundances} details the relative abundance of the initial species.

Chemistry was modelled over 1~Myr without taking dynamics and disk evolution into account, which is appropriate assumption for such chemically simple, fast-evolving species as CO, CN, and CS \citep{Semenov_Wiebe_2011}.

\subsection{Radiative Transfer}

These chemical abundances were then used for non-local thermal equilibrium (non-LTE) 3D radiative transfer modeling using \texttt{LIME} \citep{Brinch_ea_2010}. In the modelling we assumed a turbulent broadening component of $v_{\rm turb} = 0.3 c_s$ where $c_s$ is the local soundspeed \citep{Teague_ea_2016}. The collisional rates for CS were taken from \citet{Lique_ea_2010}, stored in the LAMDA database\footnote{ \url{http://home-strw.leidenuni.nl/~moldata/}}. 

A final step was to make a comparison with the observation. The \texttt{simobserve} task in \texttt{CASA} was used to simulate observations using the same array configuration in order to account for the $uv$ sampling and spatial resolution.

\section{Results}
\label{sec:results}

\subsection{Physical Structure}
\label{sec:physical_structure}

\begin{figure*}
\centering
\includegraphics[width=\textwidth]{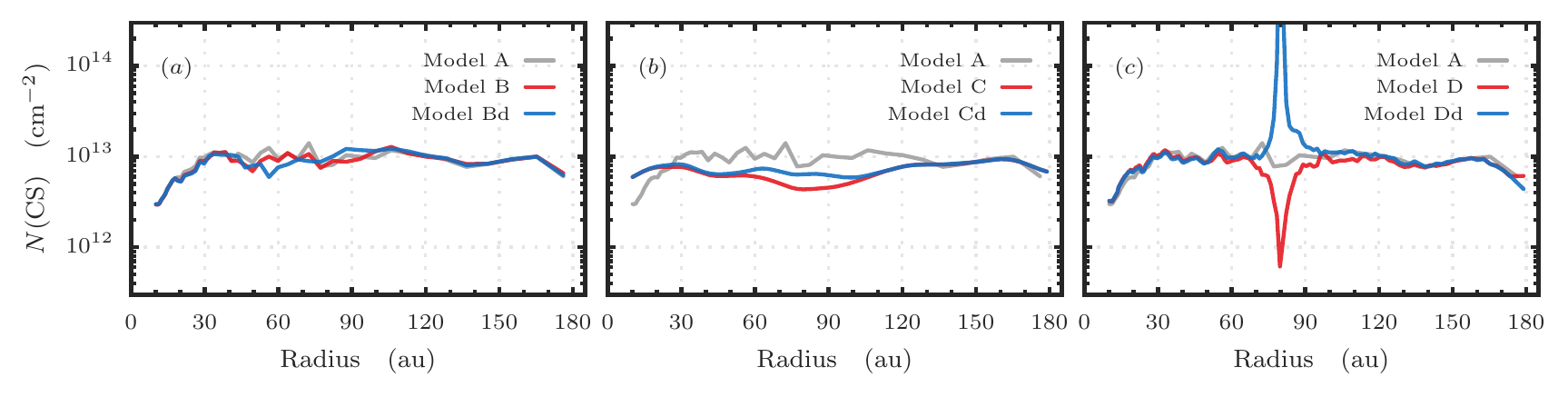}
\caption{Column densities of CS from Model A in gray, compared to those from the perturbed models. The red lines are for models with the perturbation applied to the total surface density, blue are where the perturbation is just in the dust surface density. Models B, C and D are in panels (a), (b) and (c) respectively. \label{fig:cs_columndensities}}
\end{figure*}

A depression in the surface density will create a change in the vertical structure of the disk, altering the amount of incident radiation and consequently the temperature and density of the region, impacting the chemical evolution. Figure~\ref{fig:density_temperature_slices} compares the different $n({\rm H}_2)$ and $T_{\rm gas}$ structures for the seven models around the perturbation centre. As $T_{\rm dust}$ remains largely unchanged, with average deviations from Model A of 2~\% for $z\,/\,r \leq 0.75$, we do not plot those figures. A full set of physical structure plots can be found in Appendix~\ref{sec:app:additional_figures}.

The shallow, $d = 0.3$, perturbation in Models B and Bd makes no significant change to the physical structures compared to the physical structure of the standard disk, Model A. A slight reduction of the surface density over the dip does not affect the disk temperature structure, while gas density variations over the dip are too weak to be visible in the top row of Fig.~\ref{fig:density_temperature_slices}.

With the slightly deeper perturbation in Model C, the scale height of the disk is reduced and results in a less vertically extended disk. With the depletion of smaller grains, the gas is unable to cool as efficiently in Model Cd relative to Model C, resulting in a slightly warmer region with an average temperature increase of $\approx 5$~K at the centre of the perturbation, 80~au, for $z / r \leq 0.75$.

The effects of a fully depleted region in Models D and Dd are much more prominent, as shown in the final two columns of Fig.~\ref{fig:density_temperature_slices}. For Model D, the reduction in surface density lowers the scale height of the disk, resulting in a highly irradiated outer surface. This enhances UV driven processes, such as H$_2$ pumping and photoelectric and PAH heating, resulting in a local temperature increase. In the dust depleted region in Model Dd, collisional cooling becomes less efficient, and gas gets locally hotter than in the unperturbed Model A. This results in a sharp spike in gas temperature, also increasing the gas scale height locally.

\subsection{Chemical Structure}
\label{sec:chemical_structure}

\begin{figure*}
\centering
\includegraphics{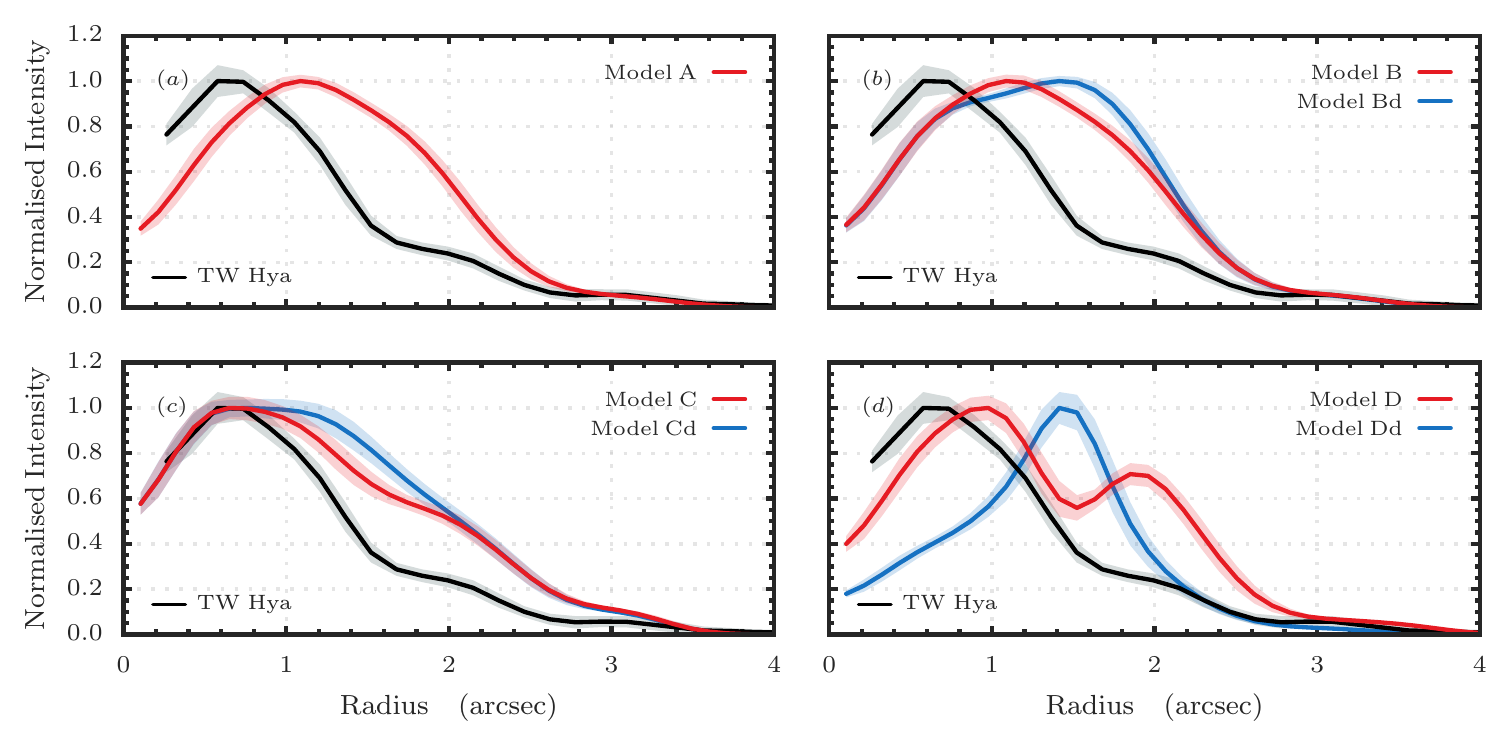}
\caption{Continuum subtracted normalised radial profiles of CS $J=5-4$ for the seven models, coloured lines, compared to TW Hya, black line. The models have been processed using the \texttt{simobserve} task in \texttt{CASA} to match the conditions of the observations. A normalised profile is used as this modelling only aims to reproduce the radial morphology of the emission profile rather than the absolute intensity. \label{fig:modelled_intensity_profiles}}
\end{figure*}

\begin{figure*}
\centering
\includegraphics[]{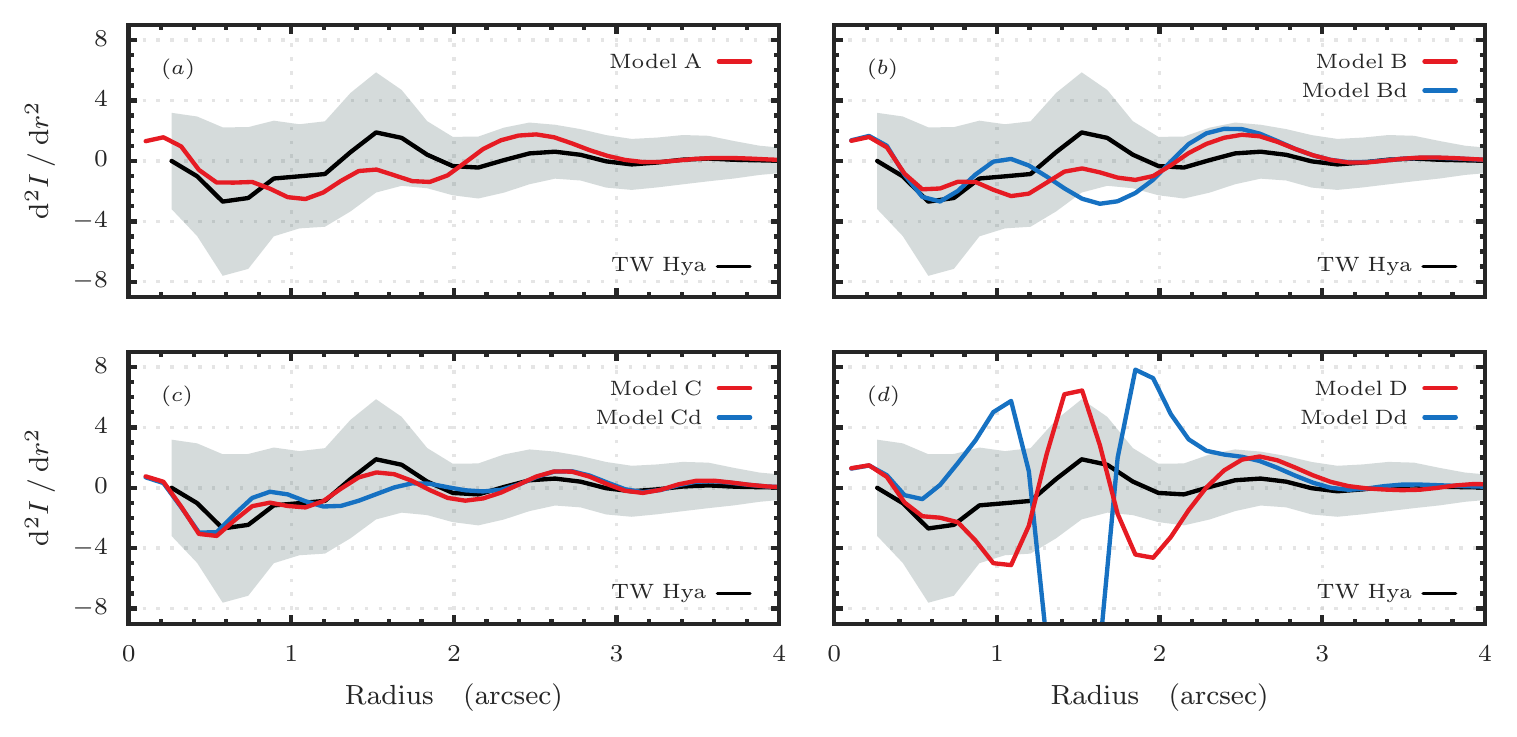}
\caption{Comparing the second derivative of the CC $J=5-4$ normalised radial intensity profiles for the seven models, coloured lines, with the observations, black line. This demonstrates that Model~C produces the most consistent radial morphology to the observations. \label{fig:modelled_second_derivatives}}
\end{figure*}

In this section we restrict our discussion to the column densities for CS. A more extensive exploration of the whole chemical inventory is left for future work. Figure~\ref{fig:cs_columndensities} shows the column densities of CS for the seven models considered. 

In the chemical network used, CS is efficiently formed through the recombination of HCS$^+$, or HOCS$^+$ at warmer temperatures, $T \gtrsim 50$~K, with H$_2$. It is primarily depleted from the gas phase by freezing out onto grain surfaces. The predicted column densities for our unperturbed Model A at $r=100$~au is $N({\rm CS}) \sim 10^{13}$~cm$^{-2}$ and are in good overall agreement with typically observed values in T~Tauri disks \citep{Guilloteau_ea_2012} and estimated for TW~Hya from single-dish observations \citep{Kastner_ea_2014}.

Models B and Bd show little change from the fiducial model, an expected outcome given the lack of change in physical structure. This demonstrates that changes in volume density on the order of 30\% are insufficient to impact the CS chemistry. In contrast, Model C shows a clear depletion in \N{CS} relative to Model A. With the reduced dust density of the disk in Models C and Cd, the harsher stellar UV radiation dominates slightly deeper into the disk, truncating the vertical extent of the CS molecular layer resulting in a narrower CS molecular layer.

The enhancement of Model Cd relative to Model C is due to the lack of grain surface onto freeze-out in the region with an enhanced gas-to-dust ratio. This reduced efficiency of depletion is more clearly seen in Model Dd which shows a spike of \N{CS} where there are no grains available for the freeze-out, in stark contrast to what is observed in TW~Hya. Furthermore, a fully depleted region of gas results in no molecular gas, as shown in Model D with a large gap in \N{CS}. Such extreme changes in column density should be readily detectable in the emission profiles.

\subsection{Molecular Emission}
\label{sec:emission}

Figure~\ref{fig:modelled_intensity_profiles} show the CS $J=5-4$ radial intensity profiles after continuum subtraction for all seven models. The left column shows the radial profile directly from LIME and the right column shows the radial profile for the simulated observations. The solid lines represent models with constant gas-to-dust ratios while dotted lines are for models with dust surface density perturbations and the shaded regions show the azimuthal variance of the radial bin. For the LIME model, the noise is intrinsic Monte-Carlo noise from the random grid in LIME.

The intensities in the inner disk are roughly a factor of 2 too low compared to the observations, although are more comparable for the outer, $r \gtrsim 100$~au, disk. For this work, we focus on the deviation between models, rather than a direct fit of the TW~Hya disk and discuss the applicability of the model to TW~Hya in Section~\ref{sec:discussion}.

As expected from the column densities, Models B and Bd exhibit no clear deviation from the unperturbed Model A, further demonstrating that a shallow dip cannot simultaneously explain the CS emission morphology and the NICMOS observations of \citet{Debes_ea_2013}. Conversely, Model D and Dd produce clear features which would be easily recognisable in the observations had the surface density perturbation been so dramatic.

Models C and Cd show a clear deficit of emission relative to Model A due to the reduced column densities described in the previous section. Both emission profiles here show a more comparable morphology to the observations than other models. This is more clearly seen in Fig.~\ref{fig:modelled_second_derivatives} which compares the normalised radial profiles with the observations. Panel (a) shows the LIME models and panel (b) shows the simulated observations. Model C clearly has a radial morphology which is comparable to the observations while Model Cd lacks the depression at $1.6\arcsec$. 

These simulated observations demonstrate that the surface density perturbation used in \citet{vanBoekel_ea_2016} is able to produce a similar radial morphology in CS $J=5-4$ emission as observed in TW~Hya. Furthermore, the more shallow perturbation used in \citet{Debes_ea_2013} was insufficient to make a noticeable change from the fiducial model and a fully depleted gap would result in features which are far more prominent in the observations.

\section{Discussion} 
\label{sec:discussion}

In the previous section we have demonstrated that a surface density perturbation, which has been found to well model the scattered light profile of TW~Hya, also produces a similar feature in the CS emission which has been observed in TW~Hya. In the following section we first discuss the applicability of the results to TW~Hya before discussing other possible mechanisms for producing the oscillatory features we observe.

\subsection{Applicability to TW~Hya}
\label{sec:applicabiliy}

Clearly these models do not fully recover the observations profiles of TW Hya: the CS $J=5-4$ line has a peak flux of 34.5~mJy~beam$^{-1}$~km\,s$^{-1}$ at a radial offset $0.58\arcsec$ while the fiducial model, Model A, has a peak flux of 26.7~mJy~beam$^{-1}$~km\,s$^{-1}$ at $1.01\arcsec$. While an iterative modelling effort was not the goal of this work, it is important to understand what may cause the discrepancy given the `TW~Hya' like fiducial model and whether this would impact the conclusions.

To verify the physical and chemical models were producing comparable flux densities to TW~Hya, radiative transfer modelling was performed for $^{12}$CO, $^{13}$CO and C$^{18}$O in Model A. Abundances for isotopologues were scaled assuming $n(^{12}{\rm CO}) \, / \, n(^{13}{\rm CO}) = 84$, consistent with the ISM \citep{Wilson_Rood_1994}, and $n(^{13}{\rm CO}) \, / \, n({\rm C^{18}O}) = 10$ which was found for TW~Hya \citep{Schwarz_ea_2016}. Flux densities were calculated for the CO $J=2-1$, $^{13}$CO $J=3-2$, $6-5$ and C$^{18}$O $J=3-2$, $6-5$ lines which all agreed within 10\% of the observed values (Schwarz, K. private communication). This gives confidence that the model used is yielding realistic abundances of the targeted molecules, despite not accurately modelling TW~Hya.

We note that while the modelled lines of CO isotopologues yield a comparable flux density, all fail to match the observed radial profile in the same manner: emission in the outer disk ($r \gtrsim 90$~au) is over produced while in the inner disk it is under produced. This is likely because the fiducial model was made to fit integrated intensities, rather than spatially resolved observations, so that the exact radial profiles in temperature and density are not matched. Nonetheless, the modelling in Section~\ref{sec:modelling} demonstrates that by perturbing an otherwise smooth surface density profile, features in the CS emission profile can be produced which are comparable in morphology to those that are observed in TW~Hya.

\subsection{An Outer Enhancement in CS}
\label{sec:outer_feature}

The observations also display a feature further out in the disk between $2.5\arcsec$ and $3.5\arcsec$, `Dip II' and `Bump II'. This feature is reproduced in all models to varying extent, as shown in Fig.~\ref{fig:modelled_second_derivatives}. Given that all models reproduce this feature, it is likely intrinsic to the set up of the models and is attributed to the exponential edge of the surface density profile used (Eqn.~\ref{eq:surfacedensity}). With the surface density dropping precipitously outside 100~au, UV radiation is able to penetrate deeper into the disk, increasing the desorption rate of volatiles.

\begin{figure}
\centering
\includegraphics[]{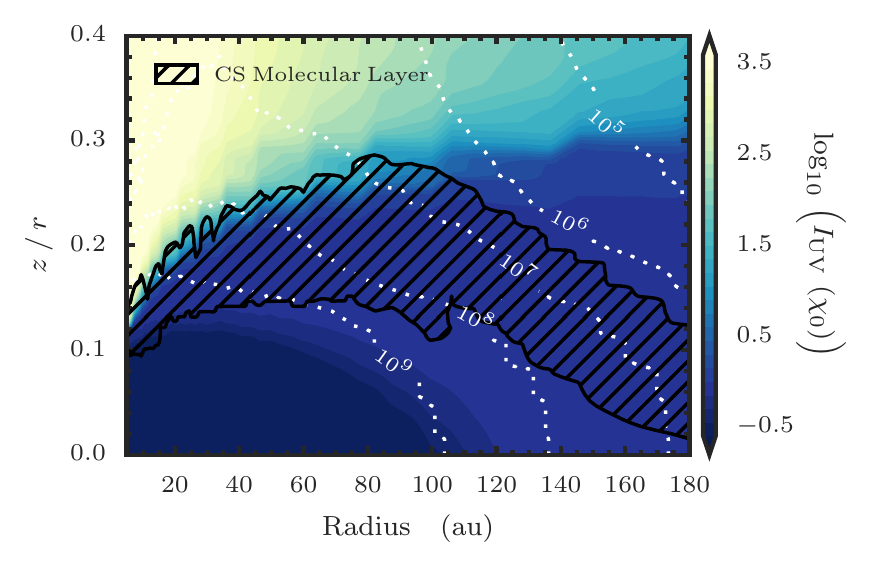}
\caption{Showing the local UV flux within Model~A in Habing units. The hatched region shows the CS molecular layer, defined as cells with contribute 1\% or more to the total column density. The white dotted contours show $n({\rm H})$ in cm$^{-3}$. \label{fig:uv_flux}}
\end{figure}

Figure~\ref{fig:uv_flux} shows the local UV flux for Model~A with the CS molecular layer, defined as cells which contribute 1\% or more to the CS column density at that radius, overlaid. Although the molecular layer traces a lower UV flux in the outer disk, the CS number density weighted average UV flux increases radially outwards from 140~au for all models. This relative enhancement in UV flux increases the intensity of photo-processing which impacts the CS abundance.

While photo-desorption of CS could directly enhance the CS abundance, other pathways are possible. Enhanced UV intensities could lead to a higher rate of dissociation of CO, releasing more C with which to form light hydrocarbons, in addition to unlocking elemental S from other more complex species. Furthermore, higher ionization rates from the increased FUV intensity could ionize S to form S$^+$, resulting in much more efficient formation of CS.

While for the chemical models we can isolate the enhanced UV intensity as the cause, potentially resulting in a photo-desorption front, a similar mechanism to that proposed by \citet{Oberg_ea_2015}, we cannot make such a claim for TW~Hya. An outer enhancement could equally be due to increased thermal desorption front \citep[for example,][]{Cleeves_2016}. To further our understanding of this feature requires a much better characterisation of the outer disk. This can be achieved with much more sensitive observations of more complex species than CO isotopologues which only extend out to $\sim 90$~au \citep{Schwarz_ea_2016}.

\subsection{Embedded Planet}
\label{sec:planets}

An attractive and commonly invoked mechanisms for explaining surface density perturbations is a planet carved gap \citep{Lin_Papaloizou_1986}. \citet{Crida_ea_2006} show that the depth and width of a depression in the surface density can be used to constrain the mass of embedded planets. \citet{Debes_ea_2013} used this relationship with their perturbed surface density to estimate an embedded planetary mass of between 6 and 28~$M_{\earth}$. 

For the partially filled case, as suggested by our modelling, \citet{Duffell_2015} derives an analytic expression for the planet mass,

\begin{equation}
q^2 = \frac{3 \pi \alpha d}{(1-d) \, f_0 \, \mathcal{M}^5},
\label{eq:planetmass}
\end{equation}

\noindent where $q = m_{\rm planet} \, / \, M_{\ast}$, $\alpha$ is the viscosity parameter \citep{Shakura_ea_1973}, $d$ is the depth as in Eqn.~\ref{eq:perturbation}, $f_0 = 0.45$, a dimensionless parameter, and $\mathcal{M} = r \, / \, H_p$ is the Mach number. Taking the depth of the perturbation from Model C, $d = 0.45$, consistent with the depth found by \citet{vanBoekel_ea_2016}, $M_{\ast} = 0.7\,M_{\sun}$ and $\alpha = 10^{-4}$ -- $10^{-3}$, values appropriate for TW~Hya \citep{Teague_ea_2016}, we find using Eqn.~\ref{eq:planetmass} a potential planet mass of 12 -- 38~$M_{\earth}$. 

A better estimation could be made with a more precisely determined depth, $d$, requiring the dip feature to be observed in multiple molecular species equally sensitive to changes in local density and constrains on the local value of $\alpha$. In addition, high-resolution observations of HCN and its isotopologues have been proposed as a chemical tracers of such an embedded planet \citep{Cleeves_ea_2015b}, allowing for a direct indication of a planet-opened gap.

\subsection{Disk Instabilities}
\label{sec:instabilities}

Pure hydrodynamical or magnetohydrodynamical instabilities can also create significant perturbations in the gas density structure \citep[see for a review][]{Turner_PPVI_2014}. These instabilities can be broadly split into two types: ones producing azimuthally symmetric features, and those producing distinct azimuthal structures.

Molecular emission from TW~Hya, in addition to the mm-continuum and scattered light observations, shows no strong azimuthal variance \citep{Qi_ea_2013, Debes_ea_2013, Kastner_ea_2015, Akiyama_ea_2015, Rapson_Kastner_ea_2015, Nomura_ea_2016, Andrews_Wilner_ea_2016, Tsukagoshi_ea_2016, Schwarz_ea_2016, vanBoekel_ea_2016}. Thus, instabilities which routinely produce vortices, spiral arms or azimuthally asymmetric gaps are likely not present at $\sim 90$~au in TW~Hya. These include the baroclinic instability \citep{Klahr_Bodenheimer_2003, Cossins_ea_2010, Douglas_ea_2013, Takahashi_ea_2016}, vertical shear instability \citep{Nelson_ea_2013, Stoll_ea_2014, Richard_ea_2016}, zombie vortex instability \citep{Marcus_ea_2015} and the Rossby wave instability \citep{Lovelace_Ea_1999, Varniere_Tagger_2006, Lyra_ea_2015}.

A more promising candidate is the magneto-rotational instability (MRI) \citep{Balbus_Hawley_1990}. \citet{Flock_ea_2015} simulated a generic T~Tauri disk assuming a gas-to-dust ratio of 100 which developed a surface density depression outside the dead-zone with a depth of $d \approx 0.5$. The change in density structure and Elsasser number over this boundary forms axisymmetric zonal flows, carving the gap. A measurement of the radial extent of the dead-zone (or confirmation of its presence) would provide a great test for this instability, however previous analyses of the ionization and turbulent velocity structures have yet to detect the edge \citep{Cleeves_ea_2015a,Teague_ea_2016}. Furthermore, \citet{Flock_ea_2015} note that no structure is seen in simulated scattered light images, contrary to the case for TW~Hya.

\subsection{Grain Evolution}

Motivated by the observed rings in scattered light, \citet{Birnstiel_ea_2015} demonstrated that grain evolution naturally leads to a region with a deficit of $\micron$~sized grains without the need to invoke perturbed surface densities. This arises because of the lack of replenishment of $\micron$~sized grains due to the slow growth and drift time-scales at this region. Such a deficit of $\micron$ sized grains manifests as a dip in scattered light emission, but not thermal mm-continuum, because of the lack of grain surface to reflect the light.

Such a feature would have implications for molecular abundances, however detailed chemical modelling of such a mechanism, requiring the transport of frozen-out volatiles, is beyond the scope of this work. Modelling of recent observations of C$_2$H and C$_3$H$_2$ in TW~Hya \citep{Bergin_ea_2016} has shown that treating grain evolution can lead to significant changes in the local abundances of volatile species and is therefore an important avenue of future exploration.

\subsection{Molecular Tracers}
\label{sec:molecules}

\begin{figure}
\centering
\includegraphics[width=\columnwidth]{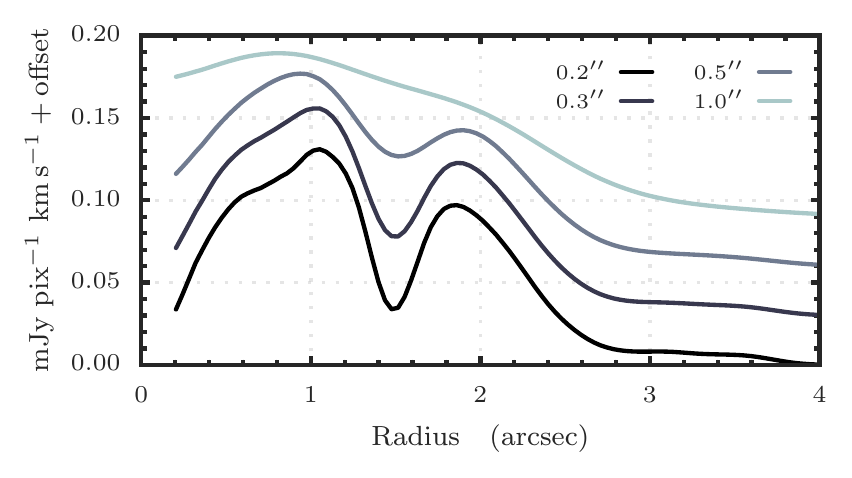} 
\caption{Radial intensity profiles for CS $J=5-4$ from Model D assuming different beam sizes. All models have the same integrated intensity. An arbitrary offset has been included to allow for easier comparison between profiles. \label{fig:beamsmear}}
\end{figure}

Despite the large archive of molecular line observations in TW~Hya, the radial feature has thus far only been observed in CS. N$_2$H$^+$ observations do not betray structure in the outer disk \citet{Qi_ea_2013}. However, the stark inner hole has been used as a proxy of the CO snowline at $\approx 30$~au which was attributed as the cause of the inner depression of $^{13}$CO and C$^{18}$O observed by \citet{Schwarz_ea_2016} and also seen by \citet{Nomura_ea_2016}. The CO isotopologue emission also showed an enhancement at $\approx 60$~au, consistent with a secondary thermal desorption front \citep{Cleeves_2016}. This suggests an enhancement in CS at $\approx 120$~au due to a secondary desorption front is an unlikely scenario due to the large distance from any clear transition in grain properties.

\citet{Kastner_ea_2015} observed C$_2$H emission in a ring between 45 and 120~au with the Sub-Millimeter Array (SMA), arguing the observed line intensities and radial location suggest a vertical segregation of grain sizes and considerable dust grain processing. \citet{Bergin_ea_2016} further argued with observations of C$_2$H and C$_3$H$_2$ that hydrocarbon rings require a local enhancement of the C\,/\,O ratio which could be achieved by the sedimentation and radial drift of grains rich in volatile ices. While these observations are not suggestive of a surface density perturbation, they do indicate that chemical abundances and the grain evolution are tightly coupled and must be considered together.
 
The modelling performed in Section~\ref{sec:modelling} suggested that molecular lines such as N$_2$H$^+$, H$_2$CO, HCN and HNC are all sensitive to small changes in the volume density, as invoked here. A full treatment of the radiative transfer for these molecules is left for future work, however it is clear that a larger repertoire of molecular tracers will provide unparalleled constraints on the physical structure of the outer disk.

\subsection{Perturbation Depth}

Better discrimination between physical mechanisms can be made with through a more thorough characterisation of the physical conditions over the gap region. As discussed above, this can be achieved through observations of molecular line emission. However, tracing features in molecular line emission rather than in continuum or scattered light will always be limited by the spatial resolution achievable. To understand how effective large beams are at masking such radial features, simulated observations of Model D were produced using range of beam sizes, $\theta_{\rm beam} = 0.2, \; 0.3,\; 0.5$ and 1$\arcsec$, as shown in Figure~\ref{fig:beamsmear}.

This demonstrates that a feature is only detectable by visual inspection for beam sizes of comparable size or smaller to the feature size. Thus, with a gap FWHM of $\approx 0.4\arcsec$ at 59~pc, only observations with spatial resolutions of $\theta_{\rm beam} \lesssim 0.5\arcsec$ would potentially detect the feature. While a partially depleted gap, such as Model C, would need to be observed at an even higher resolution in order to detect the feature. Thus it is essential for not only a range of molecular species to be observed, but also high-resolution observations in resolve the features which may be hidden with large beam sizes.

\section{Conclusion} 
\label{sec:conclusion}

We have reported observations of CS $J=5-4$ emission from the TW~Hya disk which exhibits oscillatory features outwards of $1.5\arcsec$ ($\approx 90$~au). A dip-like feature is observed at $1.6\arcsec$, comparable in morphology and position to a feature observed in the NICMOS total intensity scattered light observations \citet{Debes_ea_2013} and SPHERE polarized intensities \citep{vanBoekel_ea_2016}.

Through self-consistent thermochemical modelling and extensive chemical and radiative transfer modelling, we have demonstrated that a depression in surface density results in comparable morphology of the radial profile observed in CS at this location. A surface density perturbation with a depth of 55\% relative to a fiducial model, as used by \citet{vanBoekel_ea_2016} to model SPHERE observations of TW~Hya, produces the best fit to the morphology of the radial profile. A less severe perturbation of 30\%, used by \citet{Debes_ea_2013} to model NICMOS observations, was insufficient to strongly alter the physical and chemical structure of the disk, while a fully depleted region produced features which were far too deep. Models with the perturbation applied only to the dust surface density, therefore enhancing the local gas-to-dust ratio, were shown to impede the freeze-out of CS, resulting in an enhancement of CS emission, in clear contrast to the observations.

In addition, these models, using an exponential edge to the surface density profile, naturally reproduce the outer enhancement at $3\arcsec$. Due to the lower densities in the outer disk, inter-stellar UV can penetrate deeper into the disk. This can bolster the CS abundance through either enhanced desorption of CS, or through the unlocking of more C and S from more complex species to form CS.

These models demonstrate that both physical and chemical effects can work in tandem to produce observable features in molecular emission profiles. Only by better characterising the physical structure of a protoplanetary disk can we begin to accurately model the physical and chemical processes at play. This can be achieved by exploiting the range of molecular tracers observable with ALMA and combing these data with auxiliary data including scattered light and continuum images.

\acknowledgements
We are thankful to the anonymous referee who's comments greatly improved the quality of this manuscript. R.T. is a member of the International Max Planck Research School for Astronomy and Cosmic Physics at the University of Heidelberg, Germany. T.B. acknowledges support from the DFG through SPP 1833 “Building a Habitable Earth“ (KL 1469/13-1). D.S. acknowledges support from the Heidelberg Institute of Theoretical Studies for the project ”Chemical kinetics models and visualization tools: Bridging biology and astronomy”. This research made use of NASA’s Astrophysics Data System. This research made use of  System. This paper makes use of the following ALMA data: ADS/JAO.ALMA\#2013.1.00387.S. ALMA is a partnership of ESO (representing its member states), NSF (USA) and NINS (Japan), together with NRC (Canada), NSC and ASIAA (Taiwan), and KASI (Republic of Korea), in cooperation with the Republic of Chile. The Joint ALMA Observatory is operated by ESO, AUI/NRAO and NAOJ. This work was supported by the National Programs PCMI and PNPS from INSU-CNRS. U.G acknowledges support from NASA ADAP award NNX14AR91G. Resources supporting this work were provided by the NASA High-End Computing program through the NASA Advanced Supercomputing Division at Ames Research Center.

\bibliography{bibliography}

\newpage
\appendix
\section{Additional Figures}
\label{sec:app:additional_figures}

\begin{figure*}
\plotone{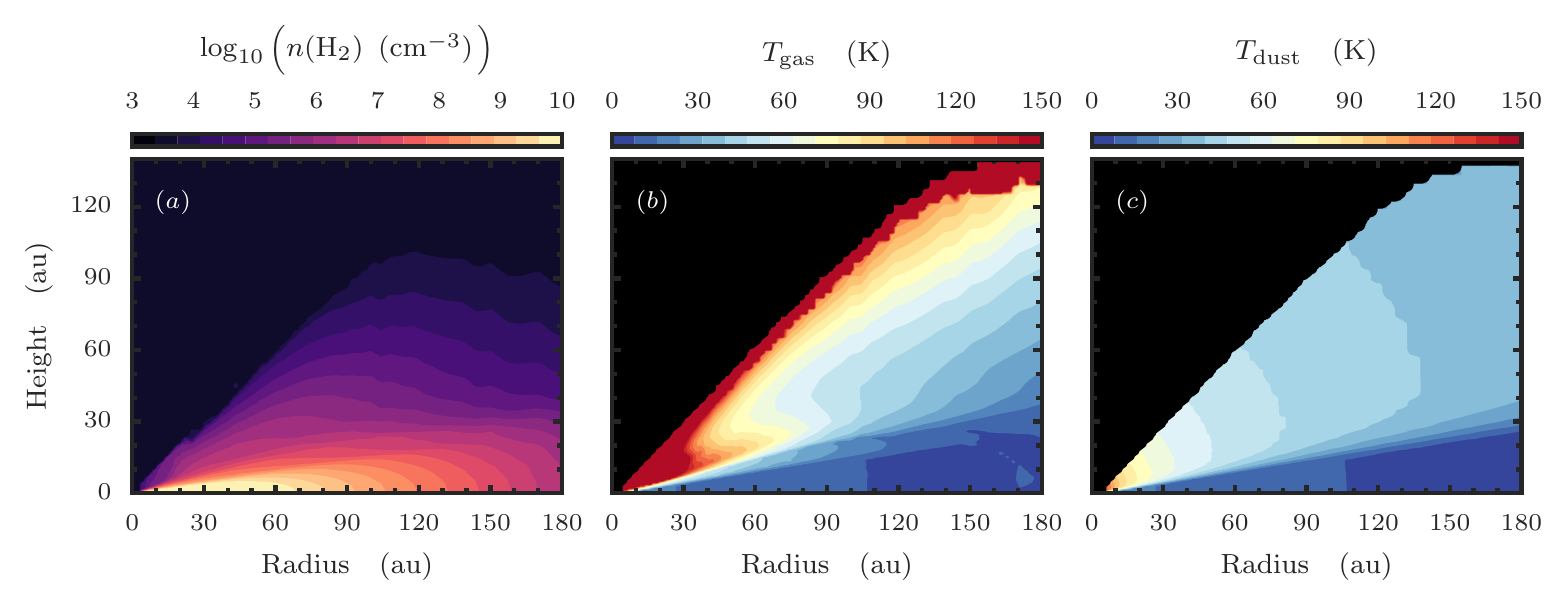}
\caption{Gas density (left), temperature (center) and dust temperature (right) for the fiducial model, Model A.}
\end{figure*}

\begin{figure*}
\plotone{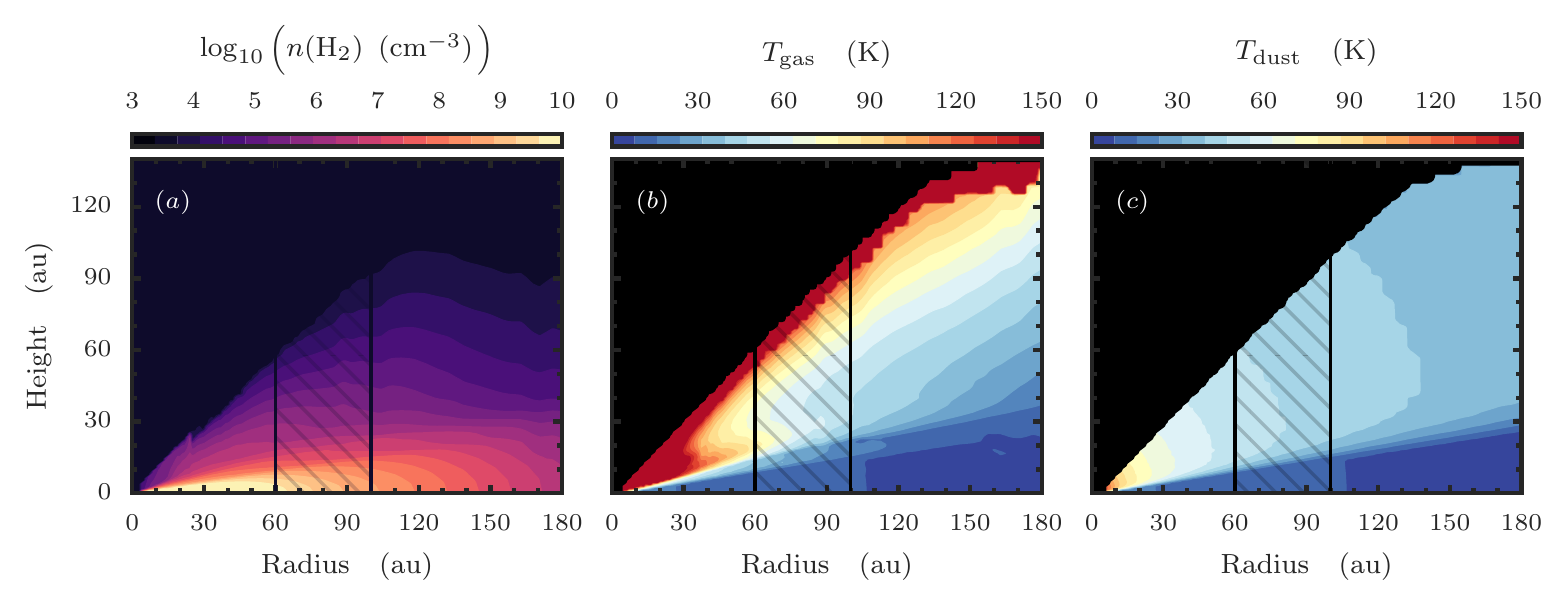}
\caption{Gas density (left), temperature (center) and dust temperature (right) for Model B.}
\end{figure*}

\begin{figure*}
\plotone{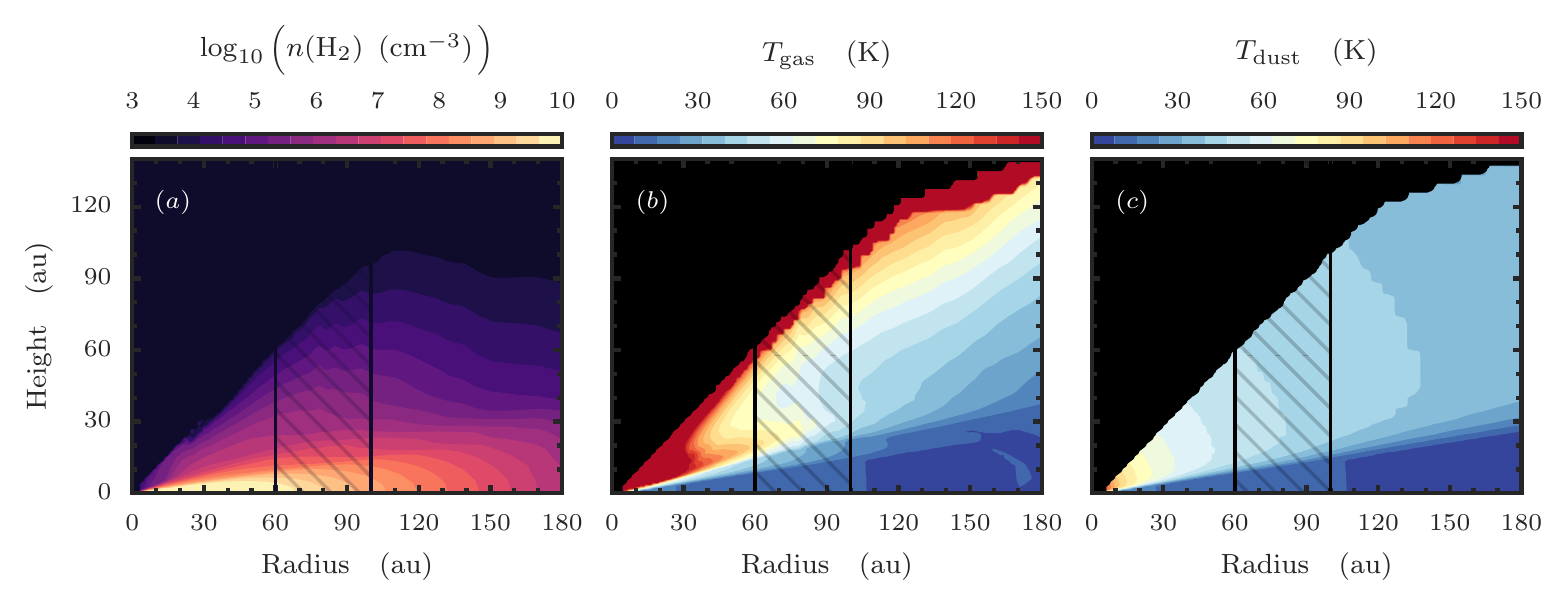}
\caption{Gas density (left), temperature (center) and dust temperature (right) for Model Bd.}
\end{figure*}

\begin{figure*}
\plotone{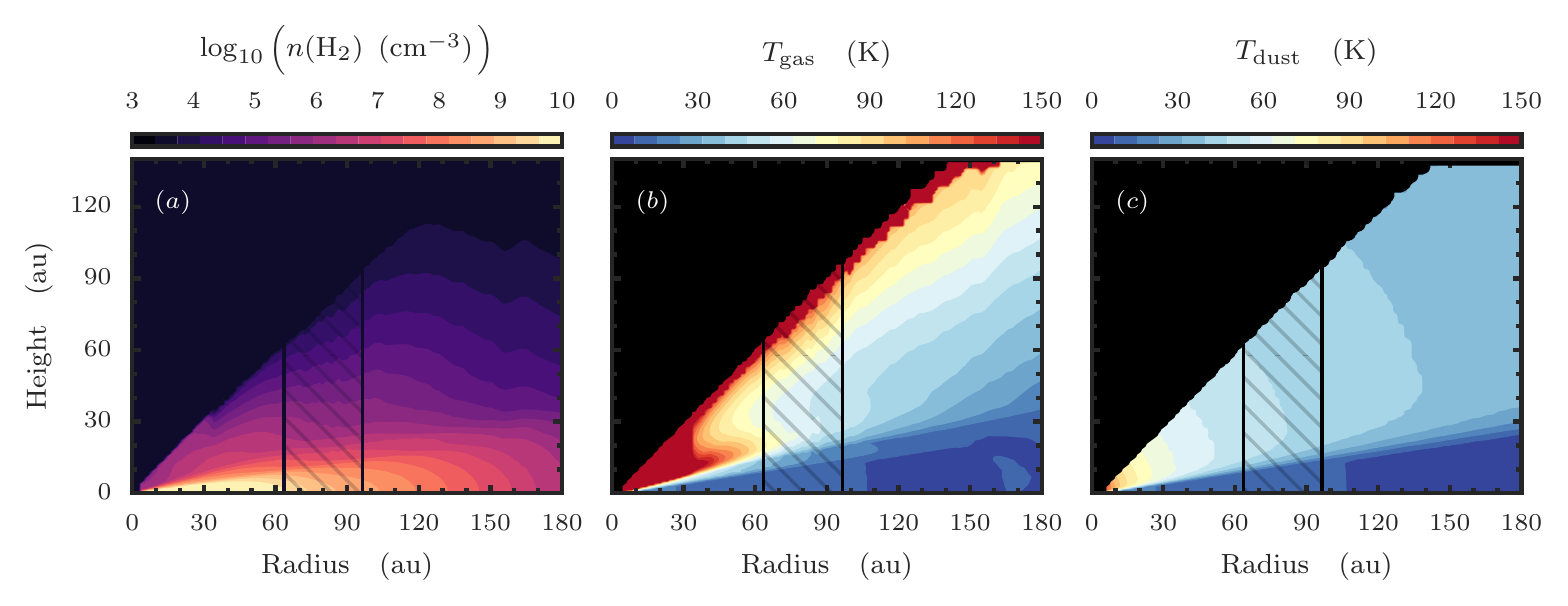}
\caption{Gas density (left), temperature (center) and dust temperature (right) for Model C.}
\end{figure*}

\begin{figure*}
\plotone{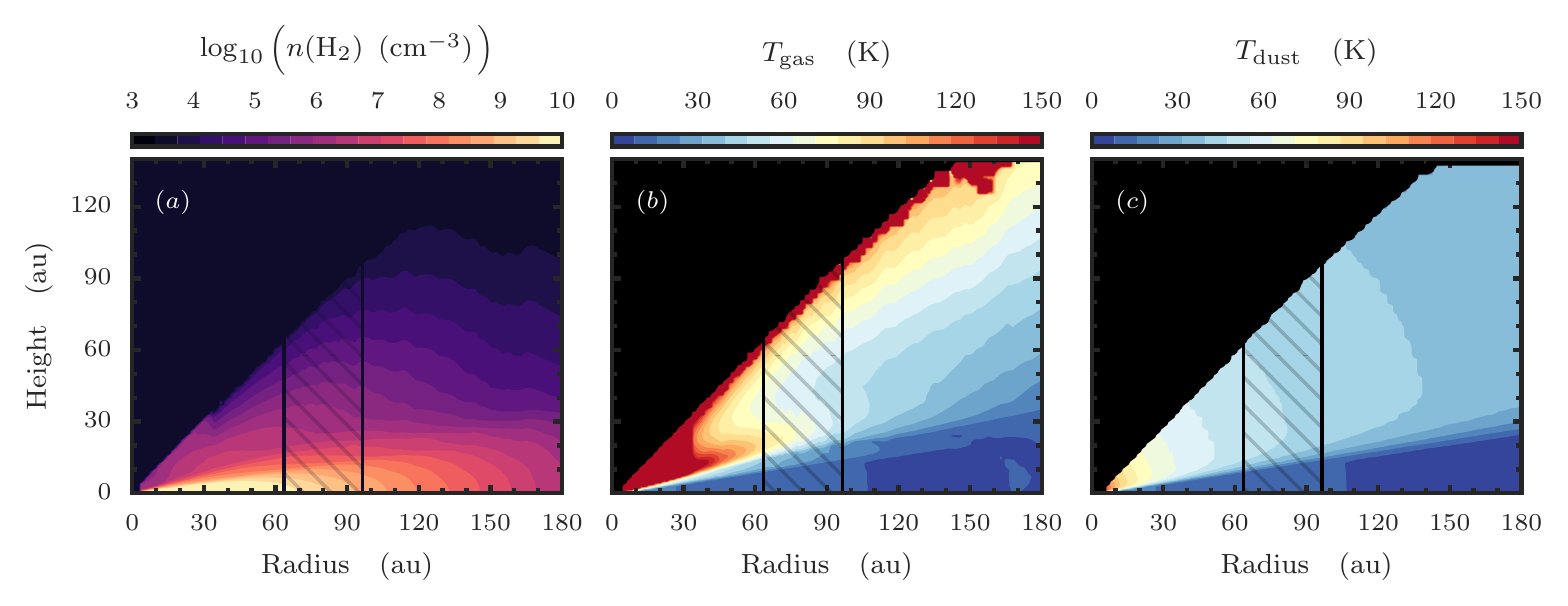}
\caption{Gas density (left), temperature (center) and dust temperature (right) for Model Cd.}
\end{figure*}

\begin{figure*}
\plotone{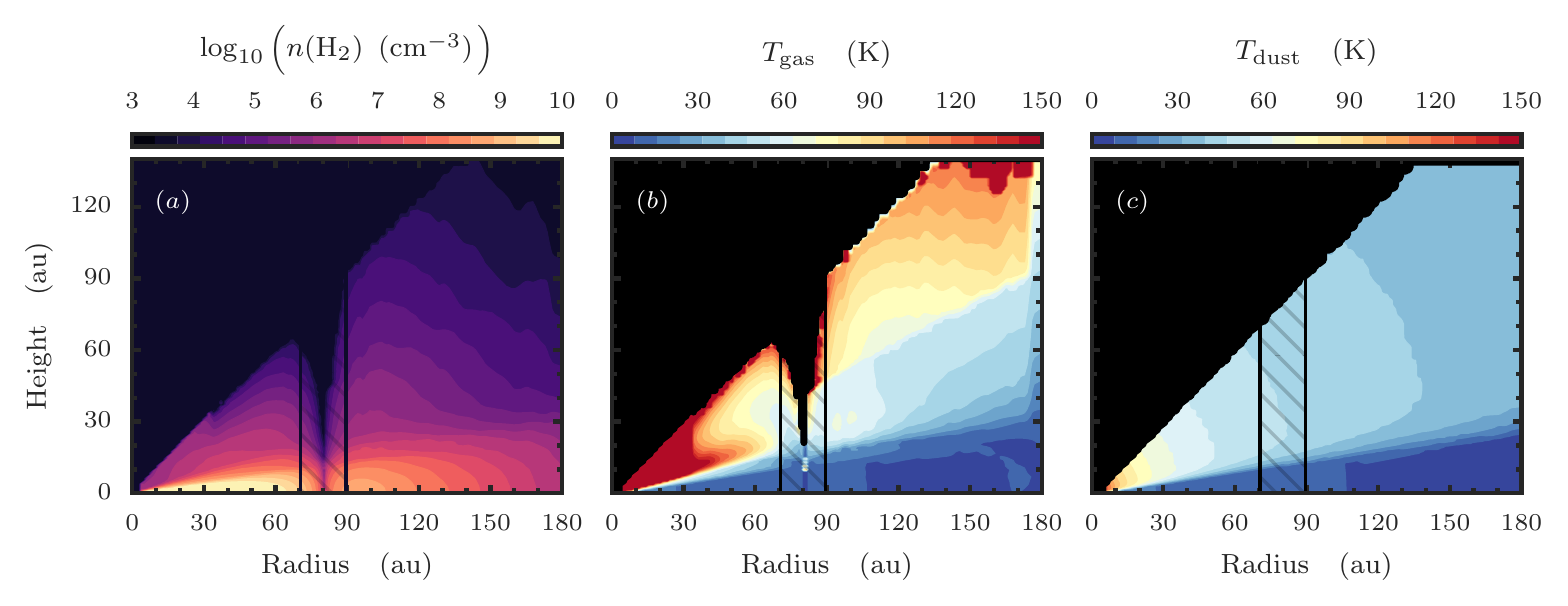}
\caption{Gas density (left), temperature (center) and dust temperature (right) for Model D.}
\end{figure*}

\begin{figure*}
\plotone{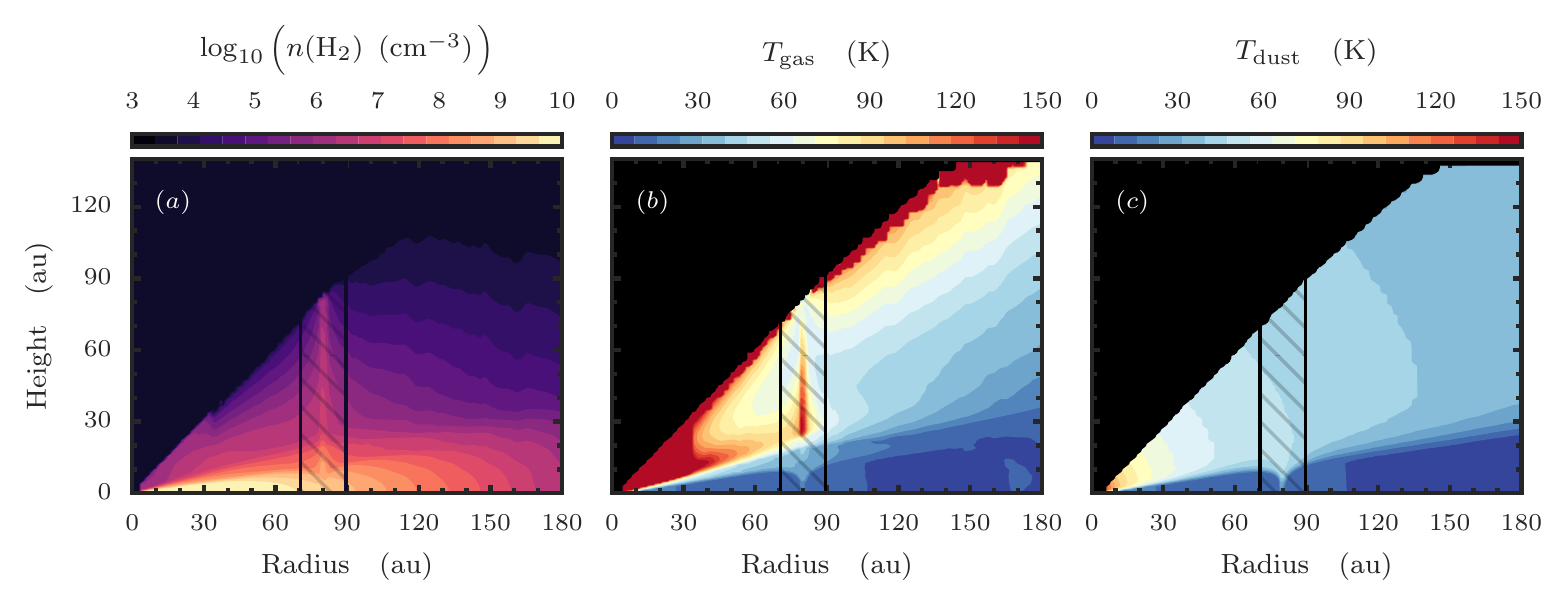}
\caption{Gas density (left), temperature (center) and dust temperature (right) for Model Dd.}
\end{figure*}

\end{document}